%% file: main.tex
\documentclass[conference]{IEEEtran}
\IEEEoverridecommandlockouts
\usepackage{amsmath,amssymb,amsfonts,subfigure,comment,eurosym,bbm}
\usepackage[]{algorithm2e}
\usepackage{algorithmic}
\usepackage{graphicx}
\usepackage{textcomp}
\usepackage{xcolor}
\usepackage{booktabs}
\usepackage{tabularx}
\usepackage{subfig}
\usepackage{flushend} 
\usepackage{lettrine}
\usepackage{url}

\IEEEpubid{\noindent 
}

\newcommand{\ind}{\mathbbm{1}}

\usepackage{acronym}

\usepackage[acronym]{glossaries}
\usepackage{glossary-mcols}
\input{acronyms}

\begin{document}

\title{A baseline Model for the Relationships between Network Operators and Tower Companies}

\author{\IEEEauthorblockN{Fabrice Guillemin and Veronica Quintuna Rodriguez}
\IEEEauthorblockA{{CNC/NARA, Orange Labs}, 
Lannion, France\\
ORCID: https://orcid.org/0000-0001-5960-2274 \& 0000-0003-0988-6242}
}

\maketitle

\begin{abstract}
The introduction of virtualization techniques in radio cellular networks allows the emergence of a business based on the outsourcing of towers hosting antennas and operated by the so-called Tower Companies (TowerCos). In this paper, we develop a baseline business model for studying the potential relationships between network operators and TowerCos. It turns out that the gain in operational costs achieved when network operators outsource the management of towers can be gracefully utilized to reduce prices so as to attract more customers. The price drop has however to be carefully realized so as not to break the market share between operators and to preserve competition. To prove this claim, we adopt in a first step a centralized optimization formulation. In a second step, we develop a game theoretic framework.
\end{abstract}

\begin{IEEEkeywords}
Virtualization, Mobile networks, TowerCos, Business model, Optimization, Game theory.
\end{IEEEkeywords}


\section{Introduction}

With the emergence of virtualization techniques and software defined networking, telecommunication networks are experiencing a major evolution of their architectures~\cite{sliceNet} and business models~\cite{campbell20175g}. \gls{NFV} \cite{ETSI_NFVArch} notably enables the decoupling between network functions and their hosting hardware; so far monolithic network functions can then be split~\cite{inriaMyPhD} and independently instantiated in data centers.

This major evolution applies to all network segments but the modifications introduced in \gls{RAN}~\cite{3GPP38_801} open the door to a complete redesign of cellular networks, for instance via the development of Cloud RAN~\cite{icin2019,HuaweiCloudRAN}, where RAN functions can be split in remote, distributed and centralized units~\cite{ORANWG1Arch}. If antennas are installed on towers equipped with a minimal amount of computing and storage resources, it becomes possible for an operator to deploy remote units on such a tower by instantiating the required software while using the resources of the tower. Operators can then envisage to outsource their towers and thus  reduce their OPEX and CAPEX~\cite{ITU2sharing,towercos2}.

This evolution perfectly fits the mobile sharing principles~\cite{GSMAsharing, ITUsharing} and notably the emerging market of Tower Companies (TowerCos)~\cite{towerco}. So far, network operators have been used to massively invest in their infrastructures, which cost several tens of  billions US dollars (USD), not only for building towers but also to connect them to the network (e.g., civil engineering, microwave or optical backhauling, etc.). Towers  actually represent a compulsory charge in mobile networks, since they are mandatory to reach mobile end users with the best possible quality of service, implying the deployment of more and more radio cells. In addition, the ever-growing demand for bandwidth and coverage as well as the proliferation of new network services \cite{5GbusinessInria} involved in smart cities, e-health, IoT, etc, require continual infrastructure upgrade and  investments in new mobile generations. 

In this context, the emergence of TowerCos introduces a new way of operating mobile networks, where  network operators invest in softwarized network functions for RAN, core, and routing networks, while TowerCos invest in infrastructures. The global telecom tower market is rapidly accelerating  with a \gls{CAGR} of almost $17\%$ (in the forecast period), which will represent an incremental growth of USD $43.71$ billion in five years~\cite{towerco}.

Currently, the main component in the value chain of a network operator is the customer basis and not the infrastructure; except for the B2B sector, the weight of services directly offered by a network operator to end users (B2C) beyond connectivity  is marginal in the global balance. Roughly speaking, while the value of a telecom infrastructure is a few billions of USD, the valorization of the customer basis, on the contrary, can be as big as several hundreds of billions USD because it opens the door to reach thousands of customers for offering added value services, mainly in an \gls{OTT} mode. 

The business of TowerCos is precisely at the confluence  of CAPEX investment by network operators and customer basis valorization by service providers. On the one hand, the outsourcing of towers allows network operators to reduce their operational costs. On the other hand, the creation of a tower business enables a rapid proliferation of towers and  thus allows more customers to be reached by service providers. It is worth noting that this is also in line with the MAGMA initiative by FaceBook~\cite{magma}, which proposes to be the intermediary between potential customers of a given area and network operators, who have no sufficient incentive to deploy a cellular infrastructure.

The emergence of TowerCos in the telecommunications market requires to find a business equation along with an equilibrium between actors with different objectives:
\begin{itemize}
    \item The TowerCo valorizes the number of customers, who can be covered by antennas installed  on  towers; this number has to be weighted by the radio quality; the achieved Quality of Service is an important factor to be taken into account in the business relationships.
    \item The network operator reduces CAPEX by outsourcing towers but increases OPEX by renting towers; the key issue for a network operator is to keep or even increase its customer basis, which is the main valorization asset.
    \item 	The objective of Service providers is to reach the largest number of end users with the best quality (for optimal Quality of Experience, QoE); the impact factor for the Service Provider in the value chain is the QoE.
\end{itemize}

To better understand the possible relationships between network operators and TowerCos, we develop in this paper a baseline business model. We consider a market with $J_0$ operators sharing a set of customers. By introducing a utility function of customers, we determine the market share between operators and notably their prices and their revenues. We then investigate what happens when operators outsource the operation of towers and use this cost reduction to decrease prices. We first formulate the problem as a centralized optimization problem, which cannot reflect the reality but helps understand the impact on the market share. We  adopt in a second step a game theoretic formulation.

This paper is organized as follows. In Section~\ref{model}, we describe the system under consideration and we introduce optimality criteria. We then evaluate in Section~\ref{value} the creation of value via the outsourcing of towers. We examine the impact on the market share when operators behave selfishly in Section~\ref{selfish} and cooperatively in Section~\ref{coordinated}. We formulate a game theoretic framework in Section~\ref{game}. Conclusions are presented in Section~\ref{conclusion}.

\section{Model description}
\label{model}
\subsection{Modeling the market share between operators}

We  develop in this section a baseline model to illustrate the role of TowerCos in the context of mobile networks. Note that we use TowerCo as a generic term to designate a variety of companies that operate towers. We first consider an area with  a tower, which is capable of covering $N$ end users. A network operator $j$ can install antennas to reach those customers; we assume that a single operator can reach up to $N$ users (this amounts to assuming that there is a strict equivalence between the deployment of a tower by an Operator or by a TowerCo). We further assume that $J_0$ operators can share the same tower. 

In a multi-operator context, customers covered by the tower have the choice between $K$ types of contracts that are described by a quality level $q =1, \ldots K$ ordered in increasing order of quality. The quality level chosen by customer $i$ is denoted by $q_i$. This quality level can for instance correspond to the amount of data that the customer can use per month in addition to the basic voice communications. The price paid by a customer of operator $j$ for quality level $q$ is denoted by $\pi_j(q)$.

The choice of an operator for a given quality level depends on the price but also on other factors, such as national coverage,  after-sale or supporting services, the quality of the hot line, etc. To reflect this situation, we introduce the utility function $u(q,j)$ for quality level $q$ by operator $j$. The utility value increases with the reputation or overall quality of operator $j$ and decreases with respect to the price (depending on the  quality level). 

In the following, we consider utility functions of the form
$$
u(q,j) = \alpha k_j - (1-\alpha)  \pi_j(q)/\pi_0
$$
where $\pi_0$ is a normalizing price constant and $\alpha$ is some positive dimensionless constant (related to user behavior and independent of the operator); this latter constant reflects the weight of the overall \textit{reputation factor $k_j$ of  operator $j$  } in the decision by a customer to subscribe a contract with this operator. The quantity $(1-\alpha)$  is the weight of \textit{the price $\pi_j(q)$} in this choice. We further assume that $k_j = a_0^j$ for some \textit{popularity index} $a_0>1$; this gives an exponential discrimination between operators, which are labeled in increased order of reputation. We thus have 
$$
u(q,j) = \alpha a_0^j - (1-\alpha)  \pi_j(q_i)/ \pi_0.
$$

With regard to prices, we further assume that they exponentially grow with the quality. This leads us to introduce a price exponent $b_j>1$ for operator $j$. In addition, an operator can be either low cost with limited offline services or else smart by offering rich services (hot line, support in shops, etc.). We assume that price exponents are ordered in increasing order of the index $j$; operator $j$ is more competitive in terms of prices than operator $j'$ if $j<j'$ (since $b_j<b_{j'}$).

With the above assumptions, the price for quality level $q$ by operator $j$ is defined as
$$
\pi_j(q) = b_j^q \pi_0
$$
for some constant price exponent $b_j>1$ depending on operator $j$ and the scaling factor $\pi_0$ (e.g., $\pi_0$ equal to 20~\euro{}).

Before proceeding to the analysis of the economic equilibrium, let us note that the system is characterized by the following  data:
\begin{enumerate}
    \item the fraction $f_q$  of users choosing quality level $q=1,\ldots, K$; 
    \item popularity index $a_0$ of operators;
    \item the parameter $\alpha$ weighting the price and the popularity of operators in the utility function of users;
    \item the price exponents $b_j$ of operators $j= 1, \ldots, J_0$.
\end{enumerate}

This baseline model assumes that depending on the price exponents $b_j$, $j=1,\ldots, J_0$, all customers choosing the quality level $q$ will subscribe a contract with the operator maximizing the utility functions $u(i,j)$ for $j=1, \ldots, J_0$ and $q(i)=q$.

We consider the revenue of operator $j$ normalized by the constant $\pi_0$ and by the number of users $N$. This quantity is defined as 
\begin{equation}
\label{defRj}
   R_j = \sum_{q=1}^K   \prod_{k=1, k\neq j}^{J_0} \ind{\{u(q,j)> u(q,k)\}} f_q\pi_j(q) ,
\end{equation}
where  $\ind_{A}= 1$ if $A$ is true and 0 otherwise. 

The fraction of users choosing operator $j$ is given by
$$
\nu_j =  \sum_{q=1}^K f_q \prod_{k=1, k\neq j}^{J_0} \ind{\{u(q,j)> u(q,k)\}}.
$$
The Average Revenue Per User (ARPU) for operator $j$ is 
$$
r_j = \frac{R_j}{ \nu_j}.
$$


\subsection{Optimality criterion}

To fix prices, network operators have to adopt a strategy to share the market. In general the smart operator $J_0$ is the incumbent operator (e.g., the historical operator in countries where the telecom market has become competitive). In addition to  make communication  services affordable to users, the regulator puts upper bounds on prices. In this paper, we assume that the regulator sets an upper bound $B$ for the price exponents $b_j$, $j=1, \ldots, J_0$. It is also worth noting that if each operator fixes its own price exponent without taking care of the others, then it is clear that the operator with the best reputation will attract all users.

To illustrate this phenomenon, we consider several criteria. Let us  denote by $R_j(b)$ the revenue of operator $j$ when the prices are $b=(b_j, j=1, \ldots, J_0)$. If we consider the global revenue of the system, equal to the sum of the revenues of the operators, we may consider the optimization problem
\begin{equation}
\label{opt0b}
\max_{b \in (1,B]^{J_0}}\sum_{j=1}^{J_0} R_j(b)
\end{equation}
 Each operator will then  try to individually optimize its own revenue by setting $b_j=B$, but the one with the best reputation will attract all customers.
 
 Now, if instead of the global revenue, we consider the criterion (known as Nash bargaining)
 \begin{equation}
\label{opt0}
\max_{b : 1<b_k<b_{J_0} <B, k=1, \ldots, J_0-1} \prod_{j=1}^{J_0} R_j(b)
\end{equation}
then numerical experiments show that customers are better distributed among operators but the situation is still unfair. For instance, for  $J_0=3$, $K=4$, $f_q=1/4$, $a_0=2$ and $B=1.5$, the normalized revenues are equal to 
\begin{equation}
R_1 = 0.0,\; R_2= 0.768208, R_3=  1.60797.
\end{equation}
We see that the operator with the worst reputation attracts no customers. This leads us to change the optimality criterion in order to achieve a fairer market share.

We still consider the product of revenues but introduce the constraint $1<b_1<b_2<\ldots < b_{J_0}<B $ so that operators fix their prices in function of their reputation. We thus consider the following optimization problem
\begin{equation}
    \label{optarpu}
    \max_{1<b_1<b_2<\ldots < b_{J_0}<B} \prod_{j=1}^{J_0} R_j(b),
\end{equation}
where $R_j$ is given by Equation~\eqref{defRj} and where the price exponents are ordered according to the reputation of the operators.

It is worth noting that the above optimization problem may still have degenerate solutions (in the sense that product is null at the optimal point $b^*$). This may occur when the coefficient $a_0$ is too small. In that case, the two cheapest operators attract all the users. The most expensive operator can attract customers only if its reputation factor is sufficiently high and the prices are not the only discriminating factors. The upper bound $B$ has hence to be carefully set in function of the reputation coefficient $a_0$.

As an illustration, we consider the case of $K=4$ quality levels and $J_0=3$ operators. There are thus a low cost operator, a smart and a medium one. The smart operator is in general the incumbent operator. Incumbent operators have in general good reputation and offer services with prices higher that the new entrants. In the numerical experiments, the weighting factor between prices and reputation is set equal to  $\alpha=.2$. We further assume that customers are evenly distributed on the 4 quality levels ($f_q=1/4$ for $q=1, \ldots,4$). We take $a_0=2$, $B=1.5$ and we solve the optimization problem~\eqref{optarpu}.

\begin{figure}[hbtp]
\centering \scalebox{.55}{\includegraphics{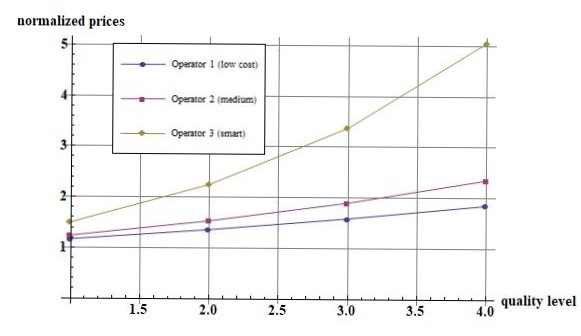}}
\caption{Normalized prices of the various operators.\label{prices0}}
\end{figure}

The normalized prices (i.e., by dropping the scaling factor $\pi_0$) are displayed in Figure~\ref{prices0}. By solving the optimization problem~\eqref{optarpu} with 
\begin{equation}
    \label{data}
    J_0=3, \;  K=4, \; f_q=1/4, \; a_0=2, \; B=1.5,
\end{equation}
we find
\begin{equation}
    \label{pricearpu}
    b_1 = 1.16635, \; b_2 = 1.23821, \; b_3=1.5
\end{equation}
The normalized revenues are equal to 
\begin{equation}
    \label{revenuesarpu}
    R_1 = 0.462657,\; R_2= 0.4746, R_3= 0.9375.
\end{equation}

\begin{figure}[hbtp]
\centering \scalebox{.47}{\includegraphics{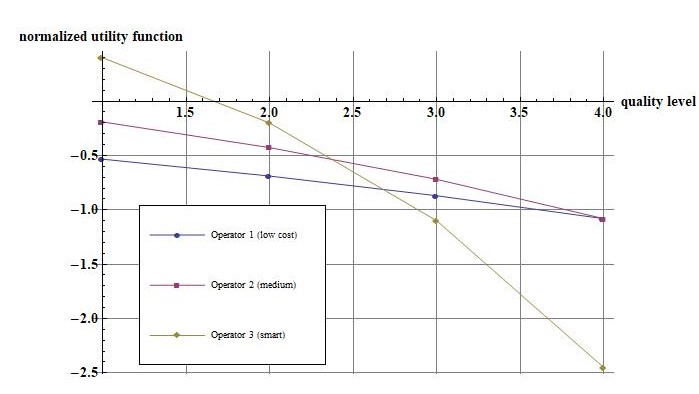}}
\caption{Utility functions of users for  the various operators.\label{utility0}}
\end{figure}

Given the prices $b_j$, $j=1, \ldots, J_0$, the utility function of the users for the various operators in function of the various quality levels are represented in Figure~\ref{utility0}. We observe that the smart operator will have users with the lowest quality levels (i.e., customers with moderated communication needs), which are ready to pay more for hot line and other offline services. The low cost operator attracts the users with the best quality because of low prices for a good quality (heavy users and rather geeks). The medium operator is in between (in general for small enterprise customers). Even if the model is oversimplified, it roughly reflects to some extent what can be observed in European markets.

With this configuration, we see that the fractions $\nu_j$, $j=1, \ldots, 4$ (i.e., the fraction of users subscribing a contract with operator $j$) are equal to 
$$
\nu_1= \nu_2=\frac{1}{4} \; \mbox{and} \; \nu_3= \frac{1}{2}.
$$
The half of the users have a contract with the smart operator and a quarter with the two other operators. The ARPUs are given by
$$
\rho_1 = 1.85063,  \; \rho_2=  1.8984, \; \mbox{and} \; \rho_3 =  1.875,
$$
and are very close to each other.

We start from this equilibrium and we analyze in the next section what happens when the operation of  towers is outsourced. We assume that this outsourcing is massive so that the revenue of operators is slightly increased by reducing the amount of OPEX and CAPEX related to the operation of towers. Globally, we assume that the revenues can be increased by a small percentage. The goal of the next section is to understand what will be the impact of this marginal increase in the revenues on the market share among the three  operators.

\section{Value creation via mutualization}
\label{value}

We assume in this section that operators massively outsource the operation of towers to TowerCo. This leads to a reduction of CAPEX and OPEX that we can pass on prices. We investigate in this section how operators can take benefit of this cost drop to lower  prices for possibly attracting more customers. 

We make the following assumptions. Let us suppose that the cost per month of the tower exploited by an  operator is $C$ (including the OPEX cost and  the CAPEX depreciation). For the sake of simplification, we assume that this cost is the same  for all operators. If the $J_0$ operators exploit their own towers, then the global  exploitation cost is $J_0 C$. Let $c$ denote the normalized price corresponding to $C$ (i.e., $C$ divided by the number of users $N$ and the scaling factor $\pi_0$). If this cost $c$ is now transferred to the TowerCo, the global normalized revenue $\mathcal{R}_j$ of  operator $j$ is increased by $c$, that is, for $j=1, \ldots, J_0$
$$
\mathcal{R}_j =  R_j +c-p_j,
$$
where $R_j$ is defined by Equation~\eqref{defRj} and  $p_j$ is the normalized  price  paid by operator $j$ to the TowerCo for using the tower. The normalized revenue of the TowerCo is moreover given by
$$
\mathcal{R}_0 = \sum_{j=1}^{J_0} p_j -c',
$$
where $c'$ is the normalized exploitation cost of the tower (OPEX and CAPEX depreciation) by the Towerco. The system is profitable for the TowerCo only if $\mathcal{R}_0 >0$.

Before outsourcing the exploitation of the tower, the global value of the system was $\sum_{j=1}^{J_0} R_j$. After outsourcing, the global value is
$$
\sum_{j=1}^{J_0} R(j) +J_0 c -c'.
$$

There is thus a value creation equal to $J_0 c -c'$, which is due to the sharing  of the tower among operators. By simply mutualizing the infrastructure, there is thus an operational gain and then a value creation. Note that we have neglected the cost of installing the antennas or other radio elements on the tower. These additional cost may be neglected when considering the installation and the operation of a tower.

To simplify the discussion, we assume in a first step that $p_j$ is the same for all operators, equal to $p$. Because of mutualization, the price $p$ should be less than the operational cost $c$, say $p = \kappa c$, with $\kappa \in (0,1)$ so that the benefit of an operator is $(1-\kappa)c$. Let us assume that the operational cost of the tower is $c'=c$ (the same as the operational cost of the tower exploited by a network operator). Under these assumptions, 
$$
\mathcal{R}_0 = (J_0 \kappa -1)c,
$$
since $\kappa$ is common to all operators. There is a benefit for the TowerCo only if $\kappa >1/J_0$

Given the potential benefit which can be achieved by  operators via the outsourcing of tower exploitation, we can envisage various scenarios. In a first one, operators can behave selfishly. In that case, an operator uses the exploitation gain to decrease its own price exponent. We investigate in the next section what happens if only one operator applies this policy. It turns out that such a selfish behavior can break the equilibrium between operators. This is why we further study the case when all operators apply this policy but in coordination. It turns out that it is possible to translate the exploitation gain into price reduction without breaking the equilibrium between operators. In practice, such a coordinated sharing between operators is not possible and we finally introduce in the subsequent section a game theoretic formulation of the problem.  

\section{Selfish behavior of operators}
\label{selfish}
Thanks to the outsourcing of the exploitation of towers,  operators increase their revenue  and can then use this marginal gain to lower prices to attract more customers and in turn  increase their global revenue. In a first step, we examine a selfish behavior by operators.

 We assume that for  operator $j$, outsourcing the exploitation of a tower increases its  normalized revenue by a factor $\varepsilon$, which is in general equal to a few percents. This operator can then lower prices by a factor $\varepsilon_j$ as long as the relative revenue reduction is smaller than $\varepsilon$.
 
Moreover, by decreasing  prices, the reputation of  operator $j$ may  also increase by a factor $\eta \varepsilon_j$ for some $\eta \in (0,1)$. In summary,  outsourcing  the operation of a tower introduces a relative gain $\varepsilon= (1-\kappa)c/R_j$ for the operator, which is a small percentage and is used to lower prices by a factor $\varepsilon_j$.
The coefficient $\eta$ is quite difficult to capture but is introduced in the following  to reflect the increase in popularity.

With all these assumptions,  operator $j$ decreases the price exponent as $(1-\varepsilon_j)b_j$ for $\varepsilon_j \in (0,1)$ and the reputation factor for this operator becomes $(1+\eta \varepsilon_j)$. The normalized utility function of customer $i$ with respect to this operator is now
\begin{equation}
    \label{utilijeps}
u(i,j) = ((1+\eta \varepsilon_j)a^j - ((1-\varepsilon_j)b_j)^{q_i}.
\end{equation}
As we consider a selfish behavior, the utility functions of other operators are left unchanged.

The price reduction coefficient $\varepsilon_j$ can be increased until the relative loss of revenue for operator $j$ exceeds $\varepsilon$. Let $R_j(\varepsilon_j)$ denote the revenue of operator $j$ for a reduction coefficient $\varepsilon_j$. Note that the initial revenue is $R_j=R_j(0)$. The reduction percentage in the revenue is
$$
\frac{R_j-R_j(\varepsilon_j)}{R_j}.
$$
Then, the maximal reduction factor $\varepsilon^*_j$ is defined by
$$
\varepsilon_j^* = \sup\left\{\varepsilon_j  : \frac{R_j-R_j(\varepsilon_j)}{R_j} < \varepsilon = \frac{(1-\kappa)c}{R_j}      \right\}.
$$

To illustrate what happens when only operator $j$ lowers  prices, we consider the case when $\varepsilon =\frac{(1-\kappa)c}{R_j}=5\%$ and $\eta= 10\%$. We progressively increases $\varepsilon_j$ up to $\varepsilon^*_j$. We start from the equilibrium presented in the previous section. 

\begin{figure}[hbtp]
\centering \scalebox{.3}{\includegraphics{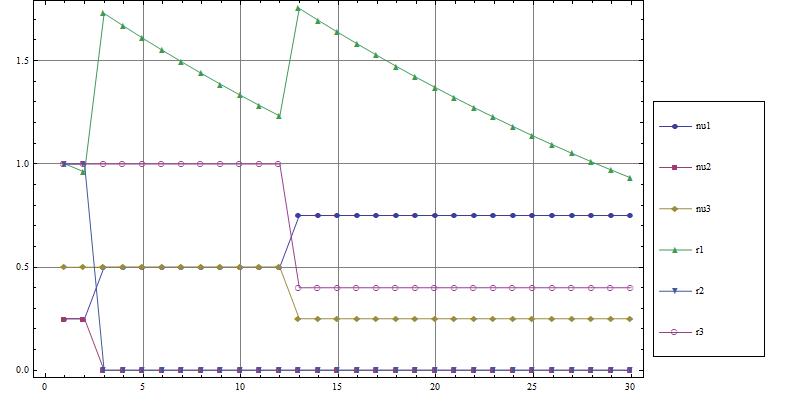}}
\caption{Operator~1 lowers price exponent $b_1$.\label{lower1}}
\end{figure}

Results for $j=1$ are presented in Figure~\ref{lower1}, i.e., when the low cost operator is the only one to lower its price exponent $b_1$.  We have represented in this figure  the faction $\nu_k$ of customers subscribing a contract with operator $k=1,2,3$ and the ratio $r_k= \frac{R_k^{(1)}(\varepsilon_j)}{R_k}$, where $R_k^{(1)}(\varepsilon_1)$ is the revenue of operator $k$ when operator 1 lowers its prices. Moreover, in this figure, the abscissa is equal to $100 \; \varepsilon_1 +1$.

\begin{figure}[hbtp]
\centering \scalebox{.5}{\includegraphics{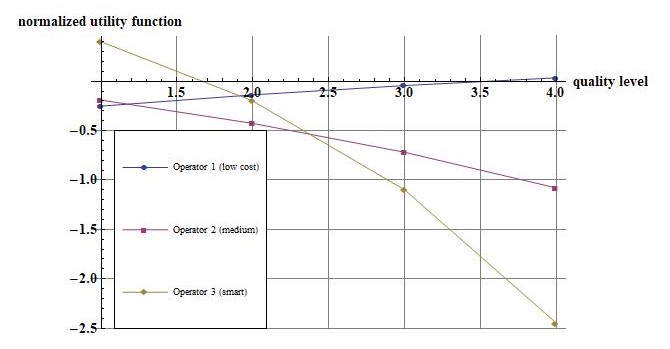}}
\caption{Utility functions when  $\varepsilon_1=\varepsilon_1^*$.\label{lower2}}
\end{figure}

We clearly see from this figure that by reducing the {price} exponent $b_1$, Operator~1 breaks the equilibrium of the previous section. Starting from a market share of 25~\% of customers, Operator~1 rapidly eliminates Operator~2 by reaching a market share equal to 50~\%. As a consequence, the corresponding revenue jumps so that $r_1=173~\%$. This gain in revenue  is used to further decrease the price exponent and gain customers to reach up to 75~\% of the market share in terms of customers. Operator~3 keeps 25~\% of the market by losing 60~\% of the initial revenue. The ultimate utility functions obtained for $\varepsilon_1=\varepsilon_1^*=29\%$ are displayed in Figure~\ref{lower2}. The smart operator keeps only the less expensive contracts \textit{(low quality level requirements)}.

\begin{figure}[hbtp]
\centering \scalebox{.4}{\includegraphics{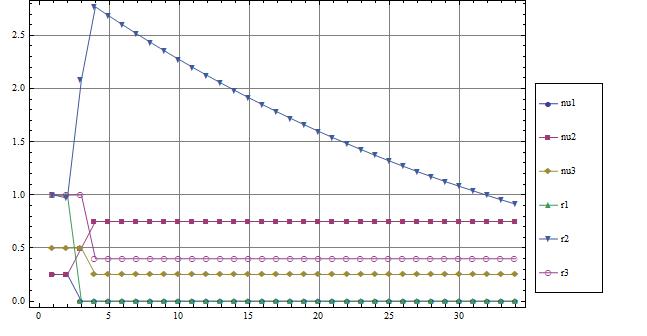}} 
\caption{Operator~2 lowers price exponent $b_2$.\label{lower3}}
\end{figure}

The same phenomenon occurs when Operator~2 is the only one to lower the price exponent $b_2$ as illustrated in Figure~\ref{lower3}. Operator~1 rapidly loses all customers. In the same way, the number of customers of Operator~3 is rapidly halved. Operator~2 can lower the price exponent $b_2$ by 33~\%.

\begin{figure}[hbtp]
\centering \scalebox{.3}{\includegraphics{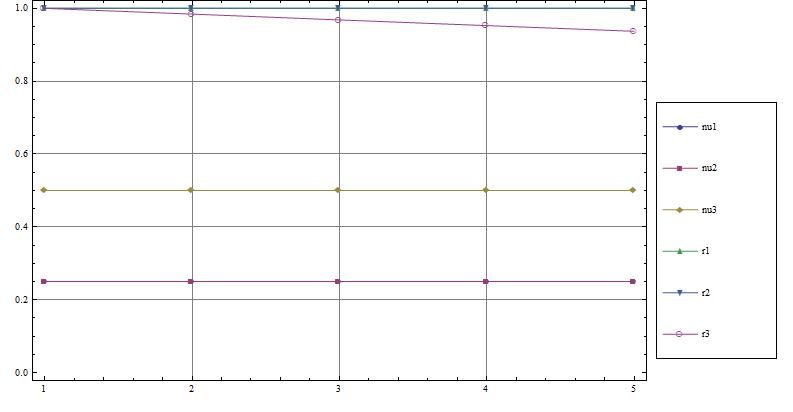}}
\caption{Operator~3 lowers price exponent $b_3$.\label{lower4}}
\end{figure}

Now, when we consider the smart operator, who has the biggest market share and the highest revenue, we see from Figure~\ref{lower4} that this operator can lower the price exponent $b_3$ up to a factor of 4~\% without losing his global revenue (the cut in revenue is compensated by outsourcing the exploitation of the tower) and without breaking the equilibrium between operators. 

We conclude from the above analysis  that the smart operator can outsource the exploitation of towers and use these savings to lower prices without breaking the market equilibrium. In fact, the loss in revenue by lowering prices is compensated by the cost reduction due to tower outsourcing. Note this reduction is allowed only because the TowerCo has a positive revenue since all  other operators outsource the tower exploitation.

It is clear that if an operator breaks the equilibrium, then the revenue of the TowerCo will decrease and the remaining operators will have to support additional charges. This phenomenon does not occur when the smart operator is the only one to lower prices. It is worth noting that lowering prices can attract more users and then increase the revenue of the operator. This is however not taken into account in the present model as this may modify the distribution of the fractions $(f_q)$ and then modify the equilibrium point. 

\section{Coordinated price reduction}
\label{coordinated}

\subsection{The tower is operated by a TowerCo}

While we have seen in the previous section that a single operator can break the equilibrium of the market share, we investigate here if it is possible that the various operators lower their respective price basis in a cooperative way without breaking the equilibrium. For this purpose, we formulate a global optimization problem. We suppose  that the outsourcing of the exploitation of the tower represents a small percentage of the normalized revenue of the smart operator, who has the largest normalized revenue. 

We assume that operator $j$ reduces its price exponent $b_j$  by a factor $\varepsilon_j$ so that the utility function is given by Equation~\eqref{utilijeps}. We set ${\epsilon} = (\varepsilon_1,\ldots,\varepsilon_{J_0})$ and we denote by $\mathbf{R}_j({\epsilon})$ the normalized revenue when the operators apply the reduction factors $\varepsilon_j$ for $j=1,\ldots,J_0$. 

We denote by $\delta$ the ratio of the gain in exploitation equal to $(1-\kappa)c$ to the normalized revenue of operator $J_0$ given by Equation~\eqref{defRj}. We assume that $\delta$ is small (say, from 1\% to 10\%), which means that the operational cost  of the tower is small when compared to the revenue of operator $J_0$. For operator $j$, the revenue $\mathbf{R}_j(\epsilon)$ after outsourcing must be such that 
$$
\mathbf{R}_j(\epsilon) + \delta R_0 > R_j.
$$
As a matter of fact, the revenue of operator $j$ after outsourcing and lowering  the price basis is  $\mathbf{R}_j(\epsilon) + \delta R_0     $ and must be beneficial for the operator so that the above inequality is satisfied. 

To determine the price lowering factors $\varepsilon_j$, we may consider a criterion which leads to a  minimization of the offset $\mathbf{R}_j(\epsilon) + \delta R_0 -R_j$ while ensuring that this quantity is positive. This leads us to introduce the quantity
\begin{multline*}
(\mathbf{R}_j(\epsilon) + \delta R_0 -R_j)^{+\infty} = \\ \left\{ \begin{array}{ll} \mathbf{R}_j(\epsilon) + \delta R_0 -R_j & if~\mathbf{R}_j(\epsilon) + \delta R_0 \geq  R_j,\\
+\infty & if~\mathbf{R}_j(\epsilon) + \delta R_0 <R_j
\end{array}
\right.
\end{multline*}
We then consider the optimization problem:
$$
\min_{\epsilon} \prod_{j=1}^{J_0} (\mathbf{R}_j(\epsilon) + \delta R_0 -R_j)^{+\infty},
$$
where $\epsilon = (\varepsilon_1, \ldots, \varepsilon_{J_0})\in [0,1]^{J_0}$. Let $\epsilon^*$ the value of $\epsilon$ at which the minimum is reached.

By solving this optimization problem with the same data as in the previous section (see Equation~\eqref{data}) and with $\delta=5\%$, we find
$$
\epsilon^* =(0.0263438, 0.0164209, 0.031625).
$$
The normalized revenues are then 
$$
\mathbf{r}(\epsilon^*) = (0.415797, 0.451602, 0.890625).
$$
When comparing the new revenues with the nominal ones, i.e., $\mathbf{R}(\epsilon)/R= (\mathbf{R}_j(\epsilon^*)/R_j, j=1, \ldots, J_0)$, we find
$$
\mathbf{R}(\epsilon^*)/R=(0.898716, 0.951542, 0.95)
$$
We see that low cost operator loses 10\% of revenue while the two others about 5\%, which is the gain due to tower operation outsourcing. But now, if we consider the global revenue $\mathcal{R}_j(\epsilon^*)$ of operator $j$ equal to $\mathbf{R}_j(\epsilon^*)+\delta R_3 e$, where $e$ is the row vector with all entries equal to 1, we have with $\mathcal{R}(\epsilon^*) = (\mathcal{R}_j(\epsilon))$
$$
\mathcal{R}(\epsilon^*) =(0.462672, 0.498477, 0.9375)
$$
so that
$$
\mathcal{R}(\epsilon^*)/R = (1.00003, 1.05031, 1.)
$$
We therefore see that there is globally no cut in the global revenues of operators, and even a slight increase for operator 2, which is of the same order of magnitude as $\delta$.

We have run the same experiment when $\delta =1\%$ and $\delta = 10\%$, we find
\begin{eqnarray*}
\epsilon^* &=& (0.00510478, 0.00146253, 0.00625461),\\
\mathbf{R}(\epsilon^*) &=& (0.453282, 0.472521, 0.92814),\\
\mathbf{R}(\epsilon^*)/R &=& (0.979737, 0.995619, 0.990016),\\
\mathcal{R}(\epsilon^*)/R & =& ( 1., 1.01537, 1.00002)
\end{eqnarray*}
and
\begin{eqnarray*}
\epsilon^* &=& (0.0550374, 0.0365904, 0.063953),\\
\mathbf{R}(\epsilon^*) &=& (0.368907, 0.424386, 0.843871),\\
\mathbf{R}(\epsilon^*)/R &=& (0.797367, 0.894196, 0.900129),\\
\mathcal{R}(\epsilon^*)/R & =& (1., 1.09173, 1.00013)
\end{eqnarray*}
respectively. We still see that the greatest loss in revenue is for the low cost operator. This is the price to pay in order to save the market share equilibrium. 

\begin{figure}[hbtp]
\centering \scalebox{.5}{\includegraphics{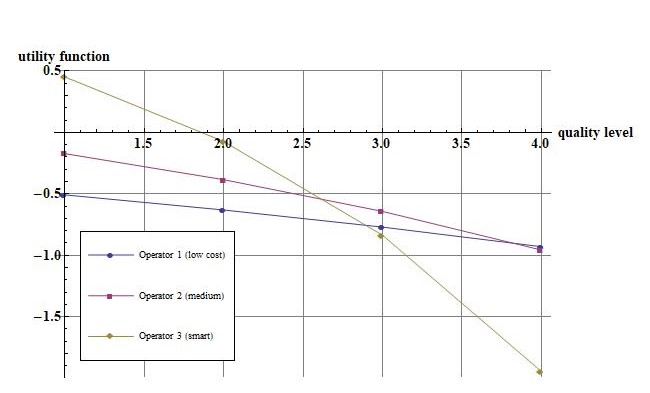}}
\caption{Utility functions of the various operators when $\delta=5\%$. \label{utilitydelta5}}
\end{figure}

The utility functions are displayed for  $\delta=5\%$ in Figure~\ref{utilitydelta5}. We clearly see from this figure the  market share equilibrium is not broken when the price exponents are cooperatively lowered by the various operators. We also empirically note that the price basis reduction is about $\delta/2$. It follows that outsourcing the operation of the tower has a positive impact on the prices paid by  customers and coordination prevents from breaking the market share between operators. Moreover, even if their "operational" revenues decrease, the global revenues are left unchanged. 

\subsection{The tower is operated by the smart operator}
Let us consider in this section the case when the smart operator operates the tower for its own usage and  the other operators. We have seen in the previous section that this is the only possibility for an operator to operate the tower since the equilibrium is rapidly broken when the low cost or medium operator slightly change their price exponents. In the case when $\delta=5\%$, we obtain
\begin{eqnarray*}
\epsilon^* &=& (0.0263522, 0.0195924, 0.031439),\\
\mathbf{R}(\epsilon^*)/R &=& (0.898685, 0.942367, 0.950291),\\
\mathcal{R}(\epsilon^*)/R & =& (1., 1.04113, 1.050293),
\end{eqnarray*}
where $\mathcal{R}(\epsilon^*) = \mathbf{R}(\epsilon^*)+\delta R_3(1,1,2)$. As a matter of fact, the smart operator still supports the operation of the tower but gets $2 \delta R_3$ paid by the other operators. 

We see that the revenues of all operators decrease but the global revenue of the low cost operator is left unchanged and the revenues of the two others increase by a percentage of the  same order of magnitude as  $\delta$. 

We have run the same experiments for $\delta$ equal to 1\% and 10\% and obtained 
\begin{eqnarray*}
\epsilon^* &=& (0.00510479, 0.00287908, 0.00611374),\\
\mathbf{R}(\epsilon^*)/R &=& (0.979737, 0.991388, 0.990241),\\
\mathcal{R}(\epsilon^*)/R & =& (1., 1.01114, 1.01024),
\end{eqnarray*}
and
\begin{eqnarray*}
\epsilon^* &=& (0.0550376, 0.0407541, 0.0618965),\\
\mathbf{R}(\epsilon^*)/R &=& (0.797367, 0.882653, 0.9032641),\\
\mathcal{R}(\epsilon^*)/R & =& (1., 1.08019, 1.10326),
\end{eqnarray*}
respectively. We observe the same phenomena as in the case $\delta=5\%$. The operators do not lose their global revenue and the medium and smart operator experience a gain of $\delta$\%.

From the above analysis, we can conclude that the smart operator can achieve some gain when operating the tower for its own purpose and for the two other operators. For the low cost  operator, this does not change its global revenue but the global revenues of the medium and smart operator are increased by a percentage $\delta$. We thus observe that the operation of the tower by the smart operator is more favorable to operators than in the case when the tower is operated by a TowerCo. The major difference is that the global value of the market is lower in the former case than in the latter. The difference in the basic situation considered in this paper is of the order of $\delta R_3$, which is used in the former situation by the smart operator to increase its global revenue while lowering its price exponent.

\section{Game theoretic formulation}
\label{game}
\subsection{Game formulation}
So far, we have considered centralized optimization by assuming that it is possible to fix the price exponents by solving a unique optimization problem. In other words, we assume that there exists an oracle capable of fixing the prices of all operators. This clearly cannot reflect the reality and we formulate in this section a distributed optimization problem. For this purpose, we consider a game composed of the $J_0$ operators and possibly the TowerCo as players.

In a first step, we examine the market share between the operators only, i.e., the operators are the only players of the game. We denote by $\mathcal{J}=\{ 1, \ldots, J_0\}$ the set of players. Let $A_j$ be the set of actions that can be taken by player $j$. In the present case, the only actions that can be taken by a player is to fix the price exponent. To formulate the game as a multiplayer matrix game, we furthermore discretize the set of possible values. Thus, $ A_j = \{\beta_{j,k}, k=1, \ldots,M_j\}$ with
\begin{equation}
    \label{defbetak}
\beta_{j,k} = 1+k \frac{B-1}{M_j},
\end{equation}
where $M_j$ is the number of possible values for the price exponent of operator $j$ and $B$ is the upper bound for price exponents. In the following, we assume that $M_j$ is the same  and hence also $A_j$ for all players, simply denoted by $A$. 

In usual Nash equilibrium problems, players have mixed policies. In the present case, this would amount to defining a set of probability distributions $\mathcal{P}_j$ on the set $A$, namely
$$
\mathcal{P}_j = \{p_j: A\to [0,1]: \sum_{a\in A } p_j(a)=1\}
 $$
The support of a strategy $p_j$  is the subset of $A$ of those elements with a positive probability (namely, $\{a \in A : p_j(a)>0\}$). In that case, the revenue of operator $j$, defined by Equation~\eqref{defRj} can be extended to account of policies as:
$$
R_j(p)= \sum_{b\in A^{J_0}} \prod_{k=1}^{J_0} p_k(b_k) R_j(b),
$$
where $R_j(b)$ is the revenue of operator $j$ for the price exponent configuration $b= (b_1, \ldots, b_{J_0})$. In this paper, we are only interested in pure policies, i.e., in policies concentrated on a single value ($p_j(b_j)=1$ for some $b_j$).

A Nash equilibrium is then a policy $p^*$ in $\mathcal{P}= \otimes_{j=1}^{J_0} \mathcal{P}_j$ such that for all $j\in \mathcal{J}$ and $a_j\in A$, 
\begin{equation}
    \label{nasheq}
    R_j(a_j,p^*_{-j})\leq R_j(p^*_j,p^*_j),
\end{equation} 
where $p^*_{-j}$ is the policy $(p^*_1, \ldots, p^*_{j-1},p^*_{j+1}, \ldots, p^*_{J_0})$.

The above Nash equilibrium problem can actually be   viewed as a multiplayer matrix game. It suffices to introduce the  $(M\times \ldots \times M)$-matrix $\mathcal{M}$ with entries for $(k_1, \ldots,k_{J_0}) \in\{1, \ldots, M\}^{J_0}$ defined by
$$
R_1(\beta_{k_1},\beta_{k_2},\ldots, \beta_{k_{J_0}}), \ldots, R_{J_0}(\beta_{k_1},\beta_{k_2},\ldots, \beta_{k_{J_0}})
$$
where the coefficients $\beta_k$ are defined by Equation~\eqref{defbetak}, where the index $j$ is ignored since the $\beta_{j,k}$ are the same for all players.

Multiplayer matrix game problems are very difficult to solve. Some algorithms are presented in \cite{nasheq}. In the present case, the game is much simpler than those considered in that reference. We consider only pure policies, which consist of fixing price exponents, and the utility functions satisfy some remarkable properties as shown below. In fact, to find a non trivial solution to the game, we shall have to introduce additional constraints.

\subsection{Resolution}

In light of the results of the previous sections, we simplify the problem by ordering the players in increasing order of price exponents (and their reputation). We thus only consider the $J_0$-tuples $(\beta_{1}, \ldots,\beta_{J_0})$ such that $1<\beta_1<\beta_2 < \ldots<\beta_{J_0}\leq B$, or equivalently with the indices $k_1<k_2<\ldots<k_{J_0}$.

\begin{figure}[hbtp]
\scalebox{.6}{\includegraphics{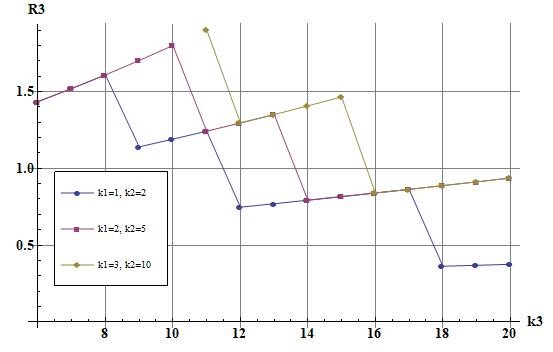}}
\caption{Values of $R_3(\beta(k_1),\beta(k_2),\beta(k_3))$ as a function of $k_3$ for different values of $(k_1,k_2)$.\label{R3k1k2k3}}
\end{figure}

In the previous section when taking $J_0=3$ operators and by setting $M=20$, we first note that for fixed $k_1$ and $k_2$ in $\{1, \ldots, M\}$ with $k_1<k_2$, the curve of $R_3(\beta(k_1), \beta(k_2), \beta(k_3))$ for $k_3= k_2+1, \ldots, M$ is piecewise increasing  and reaches a maximum for some value $K_3(k_1,k_2)$. With the same values as in the previous sections (see Equation~\ref{data}), the quantity  $R_3(\beta(k_1), \beta(k_2), \beta(k_3))$ as a function of $k_3$ for some fixed values of $k_1$ and $k_2$ is displayed for illustration in Figure~\ref{R3k1k2k3}.

\begin{figure}[hbtp]
\scalebox{.6}{\includegraphics{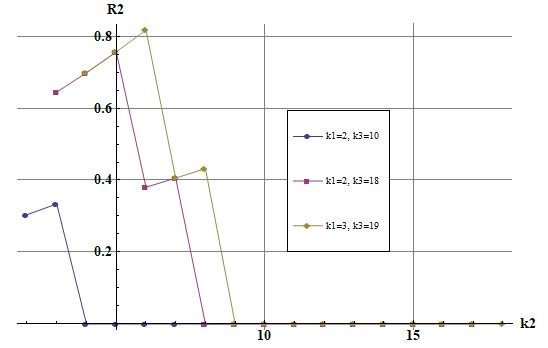}}
\caption{Values of $R_2(\beta(k_1),\beta(k_2),\beta(k_3))$ as a function of $k_2$ for different values of $(k_1,k_3)$.\label{R2k1k2k3}}
\end{figure}

The same phenomenon can be observed for  the quantity $R_1(\beta(k_1),\beta(k_2),\beta(k_3))$ (resp.  $R_2(\beta(k_1),\beta(k_2),\beta(k_3))$) when $k_1$ (resp. $k_2$) varies for fixed $(k_2,k_3)$ (resp. $(k_1,k_3)$). We have illustrated this phenomenon in Figure~\ref{R2k1k2k3} for the quantity $R_2(\beta(k_1),\beta(k_2),\beta(k_3))$ as a function of $k_2$ for some fixed values of $(k_1,k_3)$. As exhibited in this figure, the values of  $R_2(\beta(k_1),\beta(k_2),\beta(k_3))$  (and in fact also $R_1(\beta(k_1),\beta(k_2),\beta(k_3))$) are  null for a wide range of the values of $k_2$ (and $k_1$ for $R_1(\beta(k_1),\beta(k_2),\beta(k_3))$). This means that when fixing a value for $k_3$, the values of $R_1(\beta(k_1),\beta(k_2),\beta(k_3))$  and $R_2(\beta(k_1),\beta(k_2),\beta(k_3))$ are often null when the parameters $k_1$ and $k_2$ vary.

If we consider an equilibrium state $(k_1^*,k_2^*,k_3^*)$ satisfying Equation~\eqref{nasheq} for $j=3$, then $k_3^*=K_3(k_1^*,k_2^*)$, where $K_3(k_1^*,k_2^*)$ is the point at which $R_3(k_1^*,k_2^*,k_3)$ is maximum when $k_3$ varies. To find an equilibrium, we have to consider the 3-tuples $(k_1,k_2,K_3(k_1,k_2))$ with $k_1<k_2$. If we run this algorithm without any further restrictions, then we experimentally observe that the smart operator can select a price exponent, which maximizes its revenue but forces the revenues of the other operators to be null (as observed above, there is only  a restricted set of $(k_1,k_2)$ values such that $R_1(\beta(k_1),\beta(k_2),\beta(K_3(k_1,k_2)))>0$ and  $R_2(\beta(k_1),\beta(k_2),\beta(K_3(k_1,k_2)))>0$) . It follows that without further regulation, the smart operator can exclude the other operators from the game.

To remedy this situation, two additional actors can regulate the game so as to avoid the exclusion of players:
\begin{itemize}
    \item The regulation authority, which can impose rules to operators so that competition between operators is preserved and  customers have the choice among a diversity of contracts at different prices; it is worth noting that the prices by incumbents are often regulated so as to open the market.
    \item The TowerCo, whose revenue is directly proportional to the number of operators hosted by its towers and thus could deny the access to one of the operators, which applies a price policy leading to the exclusion of some others. The TowerCo then becomes a player of the game, which can indirectly influence the price policies of  operators.
\end{itemize}
For the sake of simplicity, we assume that the TowerCo charges the operators independently of their revenues. In practice, we could envisage to modulate the fee paid by an operator to install antennas on the tower as a function of its revenue or its customers basis. This policy will be investigated in further studies.

Now, we consider the case when the TowerCo becomes a player of the game and imposes to operators  the following constraint: the 3-tuple $(k_1,k_2,k_3)$ is eligible only if
\begin{multline}
    \label{condprod}
   R_1(\beta(k_1),\beta(k_2),\beta(k_3))R_2(\beta(k_1),\beta(k_2),\beta(k_3)) \\ R_3(\beta(k_1),\beta(k_2),\beta(k_3))>0. 
\end{multline}
In the search for the equilibrium state, we identify the 3-tuples $(k_1,k_2,K_3(k_1,k_2))$ such that $k_1<k_2$ and the above constraint is satisfied and which maximizes $R_3(\beta(k_1),\beta(k_2),\beta(K_3(k_1,K_2(k_1)))$; the maximum value of this quantity is reached at point $k_3^*$. This yields a set 
$$
\mathcal{L} =\{(k_1,k_2,k^*_3):  \prod_{j=1}^{J_0} R_j(\beta(k_1),\beta(k_2),\beta(k^*_3)) >0\}.
$$

The next step consists of computing the 3-tuples $(k_1,k_2,k^*_3) \in \mathcal{L}$,  that maximize  $R_2(\beta(k_1),\beta(k_2),\beta(k_3^*))$, which is reached at point $k_2^*$. This gives rise another set
$$
\mathcal{L}' =\{(k_1,k^*_2,k^*_3):(k_1,k_2,k_2) \in \mathcal{L}\}.
$$
The final step is to fix $k_1^*$ realizing the maximum value of $R_1(\beta(k_1),\beta(k^*_2),\beta(k_3^*)$. 

By applying the above algorithm for $M=50$, we find
$k_1^* = 27$, $ k_2^*= 33$ and $k_3^* = 50$ with price exponents $\beta(k_1^*) = 1.26$, $\beta(k_2^*) = 1.33$, and $ \beta(k_3^*) = 1.5$; the revenues are
$$
R_1^*=0.650362, R_2^*= 0.588159,  R_3^*= 0.9375.
$$
We observe that the revenues resulting from the Nash equilibrium are slightly different from those obtained by solving the optimization problem~\eqref{optarpu}. The revenue of the smart operator is left unchanged, the revenue of the low cost is larger and even higher to that of the medium one. 

\section{Conclusion}
\label{conclusion}

We have introduced in this paper a baseline model to illustrate the market share between operators with various levels of reputation. We have further considered that the prices by an operator are geometric in terms of the level of quality (definition of a price exponent). By defining utility functions of customers, we have introduced various optimality criteria to fix prices. By ordering price exponents in terms  of reputation levels, we have been able to define an optimality criterion, which allows a market share between operators without excluding any of them. This leads to the definition of an equilibrium (see Problem~\eqref{optarpu}).

We have subsequently examined the situation when operators can achieve some gain by delegating the operation of towers hosting their antennas to a TowerCo (the generic name of companies owning towers). We have investigated the impact of this outsourcing procedure on the equilibrium, when operators behave selfishly or cooperatively. It turns out that a selfish behavior can rapidly break the equilibrium of the market share. A cooperative behavior can gracefully lower prices. 

We have then formulated a game theory problem to study the evolution of the market share. It turns out that it is necessary to introduce some rules in order to avoid the exclusion of a player (i.e., a network operator). Such rules can be imposed by the regulation authority or the TowerCo, whose revenue is proportional to the number of players.

\bibliographystyle{IEEEtran}
\bibliography{biblio}

\flushend
\pagebreak

\end{document}

%% file: acronyms.tex
\newacronym{AAA}{AAA}{Authentication, Authorization, and Accounting}
\newacronym{RSC}{RSC}{Recursive  Systematic Convolutional}
\newacronym{LLR}{LLR}{Log-Likelihood Ratio}
\newacronym{FS}{FS}{Functional Split}
\newacronym{BBU}{BBU}{Base Band Unit}
\newacronym{COTS}{COTS}{Commercial off-the-shelf }
\newacronym{VNF}{VNF}{Virtualized Network Function}
\newacronym{VNF FG}{VNF FG}{VNF Forwarding Graph}
\newacronym{NFV}{NFV}{Network Function Virtualization}
\newacronym{GPP}{GPP}{General Purpose Processor}
\newacronym{vEPC}{vEPC}{virtual Evolved Packet Core}
\newacronym{LTE}{LTE}{Long Term Evolution}
\newacronym{uRLLC}{uRLLC}{Ultra-Reliable Low-Latency Communications}
\newacronym{eMBB}{eMBB}{enhanced Mobile BroadBand}
\newacronym{mMTC}{mMTC}{massive Machine Type Communications}
\newacronym{OSM}{OSM}{Open Source MANO}
\newacronym{C-EPC}{C-EPC}{Cloud-EPC}
\newacronym{EPCaaS}{EPCaaS}{EPC as a Service}
\newacronym{TDD}{TDD}{Time Division Duplex}
\newacronym{UE}{UE}{User Equipment}
\newacronym{HARQ}{HARQ}{Hybrid Automatic Repeat-Request}
\newacronym{PRB}{PRB}{Physical Resource Blocks}
\newacronym{MCS}{MCS}{Modulation and Coding Scheme}
\newacronym{CQI}{CQI}{Channel Quality Indicator}
\newacronym{DC}{DC}{Dedicated Core}
\newacronym{RR}{RR}{Round Robin}
\newacronym{G}{G}{Greedy}
\newacronym{VPN}{VPN}{Virtual Private Network}
\newacronym{MPLS}{MPLS}{Multiprotocol Label Switching}
\newacronym{OWL}{OWL}{Web Ontology Language}
\newacronym{NST}{NST}{Network Slice Template}
\newacronym{NSST}{NSST}{Network Slice Subnet Template}
\newacronym{NSMF}{NSMF}{Network Slice Management Function}
\newacronym{NSSMF}{NSSMF}{Network Slice Subnet Management Function}
\newacronym{CSMF}{CSMF}{Communication Service Management Function}
\newacronym{FCAPS}{FCAPS}{Fault-management Configuration Accounting Performance and Security}
\newacronym{RF}{RF}{Radio Frequency}
\newacronym{RIC}{RIC}{RAN Intelligent Controller}

\newacronym{CNF}{CNF}{Cloud-Native Network Function}
\newacronym{PLMN}{PLMN}{Public Land Mobile Network}
\newacronym{CLAMP}{CLAMP}{Closed Loop Automation Management Platform}

\newacronym{FDD}{FDD}{Frequency Division Duplex}
\newacronym{OFDM}{OFDM}{Orthogonal Frequency Division Multiplexing}

\newacronym{VM}{VM}{Virtual Machine}
\newacronym{PDCP}{PDCP}{Packet Data Convergence Protocol}
\newacronym{MAC}{MAC}{Medium Access Control}
\newacronym{RLC}{RLC}{Radio Link Control}
\newacronym{RRC}{RRC}{Radio Resource Control}
\newacronym{AM}{AM}{Acknowledged Mode}
\newacronym{UM}{UM}{Unacknowledged Mode}
\newacronym{TM}{TM}{Transparent Mode}
\newacronym{MIMO}{MIMO}{Multiple Input Multiple Output}
\newacronym{MISO}{MISO}{Multiple Input Single Output}
\newacronym{SIMO}{SIMO}{Single Input Multiple Output}
\newacronym{SISO}{SISO}{Single Input Single Output}
\newacronym{MCC}{MCC}{Mobile Country Code}
\newacronym{MNC}{MNC}{Mobile Network Code}
\newacronym{S-TMSI}{S-TMSI}{Shortened Temporary Mobile Subscriber Identity}
\newacronym{IMSI}{IMSI}{International Mobile Subscriber Identity}
\newacronym{DRB}{DRB}{Dedicated Radio Bearer}
\newacronym{GUMMEI}{GUMMEI}{Globally Unique MME Identity}

\newacronym{PCI}{PCI}{Physical-layer Cell Identity}
\newacronym{ROHC}{ROHC}{Robust Header Compression}
\newacronym{SN}{SN}{Sequence Number}
\newacronym{RAR}{RAR}{Random Access Response}
\newacronym{C-RNTI}{C-RNTI}{Cell Radio Network Temporary Identifier}
\newacronym{BSR}{BSR}{Buffer Status Report}
\newacronym{DRX}{DRX}{Discontinuous Reception}
\newacronym{PHR}{PHR}{Power Head Room}
\newacronym{PUSCH}{PUSCH}{Physical Uplink Shared Channel}
\newacronym{ADM}{ADM}{Activation/Deactivation MAC}
\newacronym{GP}{GP}{Gap Period}

\newacronym{RE}{RE}{Resource Element}
\newacronym{RB}{RB}{Resource Block}
\newacronym{REG}{REG}{Resource Element Group}
\newacronym{CSRS}{CSRS}{Cell-Specific Reference Signal}
\newacronym{IFFT}{IFFT}{Inverse Fast Fourier Transform}
\newacronym{OFDMA}{OFDMA}{Orthogonal Frequency Division Multimple Access}
\newacronym{CRC}{CRC}{Cyclic Redundancy Check}
\newacronym{SFC}{SFC}{Service Function Chain}

\newacronym{eNB}{eNB}{Evolved NodeB}
\newacronym{RAN}{RAN}{Radio Access Network}
\newacronym{ARQ}{ARQ}{Automatic Repeat reQuest}
\newacronym{NAS}{NAS}{Non-Access Stratum}
\newacronym{MME}{MME}{Mobility Management Entity}
\newacronym{MIB}{MIB}{Master Information Block}
\newacronym{SIB}{SIB}{System Information Block}
\newacronym{RSRP}{RSRP}{Reference Signal Received Power}
\newacronym{RAT}{RAT}{Radio Access Technologie}
\newacronym{ACK}{ACK}{Acknowledge}
\newacronym{NACK}{NACK}{Negative acknowledge}
\newacronym{PDCCH}{PDCCH}{Physical Downlink Control Channel}
\newacronym{SAW}{SAW}{Stop and Wait}
\newacronym{TTI}{TTI}{Transmission Time Interval}
\newacronym{RRH}{RRH}{Radio Remote Head}
\newacronym{SNIR}{SNIR}{Signal-to-Noise-plus-Interference Ratio}
\newacronym{WCET}{WCET}{Worst Case Execution Time}
\newacronym{GPC}{GPC}{General Purpose Computer}
\newacronym{KPI}{KPI}{Key Performance Indicator}
\newacronym{OAI}{OAI}{Open Air Interface}
\newacronym{IMS}{IMS}{IP Multimedia Subsystem}
\newacronym{vIMS}{vIMS}{virtual IP Multimedia Subsystem}
\newacronym{EPC}{EPC}{Evolved Packet Core}
\newacronym{SDN}{SDN}{Software Defined Network}
\newacronym{C-RAN}{C-RAN}{Centralized-RAN}
\newacronym{OS}{OS}{Operating System}
\newacronym{TB}{TB}{Transport Block}
\newacronym{TBS}{TBS}{Transport Block Size}
\newacronym{QCI}{QCI}{QoS Channel Indicator}

\newacronym{GPU}{GPU}{Graphics Processing Unit}
\newacronym{CPU}{CPU}{Central Processing Unit}

\newacronym{SDU}{SDU}{Service Data Unit}
\newacronym{CBS}{CBS}{Code Block Size}
\newacronym{CB}{CB}{Code Block}
\newacronym{SPMD}{SPMD}{Single Program Multiple Data}
\newacronym{SIMD}{SIMD}{Single Instruction Multiple Data} 

\newacronym{SINR}{SINR}{Signal-to Interference Noise Ratio}
\newacronym{CO}{CO}{Central Office}
\newacronym{CA}{CA}{Carrier Aggregation}
\newacronym{SRS}{SRS}{Sound Reference Signal}
\newacronym{SC-OFDMA}{SC-OFDMA}{Single Carrier - Orthogonal Frequency Division Multiple Access}
\newacronym{FPGA}{FPGA}{Field-Programmable Gate Array}
\newacronym{TA}{TA}{Time Advancing}
\newacronym{CoMP}{CoMP}{Coordinated Multi-point}
\newacronym{NPRB}{NPRB}{Number of Physical Resource Blocks}
\newacronym{RTT}{RTT}{Round Trip Time}
\newacronym{CPRI}{CPRI}{Common Public Radio Interface}
\newacronym{CBR}{CBR}{Constant Bit Rate}
\newacronym{NRB}{NRB}{Number of Resource Blocks}
\newacronym{BJF}{BJF}{Biggest Job First}
\newacronym{EDF}{EDF}{Earliest Deadline First}
\newacronym{FCFS}{FCFS}{First-come, First-served}
\newacronym{PSTN}{PSTN}{Public Switched Telephone Network}
\newacronym{ETSI}{ETSI}{European Telecommunications Standards Institute}
\newacronym{vBBU}{vBBU}{virtualized BBU}
\newacronym{vRAN}{vRAN}{virtualized RAN}
\newacronym{IoT}{IoT}{Internet of Things}
\newacronym{B2B}{B2B}{Business to Business}
\newacronym{B2C}{B2C}{Business to Customer}
\newacronym{QoE}{QoE}{Quality of Experience}
\newacronym{QoS}{QoS}{Quality of Service}
\newacronym{VNO}{VNO}{Virtual mobile Network Operator}
\newacronym{SLA}{SLA}{Service Level Agreement}
\newacronym{VRRM}{VRRM}{Virtual Radio Resource Management}
\newacronym{KVM}{KVM}{Kernel-based Virtual Machine}
\newacronym{LXC}{LXC}{Linux Containers}
\newacronym{PS}{PS}{Processor Sharing}

\newacronym{eCPRI}{eCPRI}{evolved CPRI}
\newacronym{RoE}{RoE}{Radio over Ethernet}
\newacronym{PAPR}{PAPR}{Peak-to-average power ratio}
\newacronym{SC-FDMA}{SC-FDMA}{Single Carrier Frequency Division Multiple Access}
\newacronym{AGC}{AGC}{Automatic Gain Control}
\newacronym{PMD}{PMD}{Polarization Mode Dispersion}
\newacronym{ADC}{ADC}{Analogic-Digital Converter}

\newacronym{IQ}{IQ}{In-Phase Quadrature}
\newacronym{xRAN}{xRAN}{extensible Radio Access Network}
\newacronym{ISI}{ISI}{Inter-symbol interference}
\newacronym{FFT}{FFT}{Fast Fourier Transform}
\newacronym{IPC}{IPC}{Inter process communication}
\newacronym{CCDU}{CCDU}{Channel Coding Data Unit}
\newacronym{CC}{CC}{Channel Coding}
\newacronym{gNB}{gNB}{next-Generation Node B}
\newacronym{EUTRAN}{EUTRAN}{Evolved Universal Terrestrial Radio Access Network}
\newacronym{SCTP}{SCTP}{Stream Control Transmission Protocol}
\newacronym{NR}{NR}{New Radio}
\newacronym{NF}{NF}{Network Function}
\newacronym{CU}{CU}{Central Unit}
\newacronym{DU}{DU}{Distributed Unit}
\newacronym{NGC}{NGC}{Next Generation Core}
\newacronym{DL}{DL}{down-link}
\newacronym{UL}{UL}{up-link}
\newacronym{LJF}{LJF}{Largest Job First}
\newacronym{RANaaS}{RANaaS}{RAN as a Service}
\newacronym{NaaS}{NaaS}{Network as a Service}
\newacronym{NS}{NS}{Network Service}
\newacronym{FG}{FG}{Forwarding Graph}
\newacronym{VNFC}{VNFC}{VNF Component}
\newacronym{MANO}{MANO}{Management and Orchestration}
\newacronym{FIFO}{FIFO}{First In Firs Out}
\newacronym{NFVI}{NFVI}{NFV Infrastructure}
\newacronym{NFVO}{NFVO}{NFV Orchestrator}
\newacronym{PoP}{PoP}{Point of Presence}
\newacronym{NAT}{NAT}{Network Address Translation}
\newacronym{CDN}{CDN}{Content Delivery Network}
\newacronym{VNFM}{VNFM}{VNF Manager}
\newacronym{EM}{EM}{Element Management}
\newacronym{VIM}{VIM}{Virtualised Infrastructure Manager}

\newacronym{e2e}{e2e}{end-to-end}

\newacronym{AMF}{AMF}{Access and Mobility Management Function}
\newacronym{SMF}{SMF}{Session Management Function}
\newacronym{UPF}{UPF}{User Plane Function}
\newacronym{PCF}{PCF}{Policy Control Function}
\newacronym{UDM}{UDM}{Unified Data Management}
\newacronym{NRF}{NRF}{NF Repository Function}
\newacronym{AUSF}{AUSF}{Authentication Server Function}
\newacronym{API}{API}{Application Programming Interface}
\newacronym{HSS}{HSS}{Home Subscriber Server}
\newacronym{PCRF}{PCRF}{Policy and Charging Rules Function}
\newacronym{SOA}{SOA}{Software-Oriented Architecture}
\newacronym{AKA}{AKA}{Authentication and Key Agreement}
\newacronym{AF}{AF}{Application Function}
\newacronym{NEF}{NEF}{Network Exposure Function}
\newacronym{NSSF}{NSSF}{Network Slice Selection Function}
\newacronym{NSSP}{NSSP}{Network Slice Service Profile}
\newacronym{VES}{VES}{Virtual Event Streaming}
\newacronym{NSSAI}{NSSAI}{Network Slice Selection Assistance Information}

\newacronym{NSSI}{NSSI}{Network Slice Subnet Instance}

\newacronym{NSS}{NSS}{Network Slice Subnet}
\newacronym{NSC}{NSC}{Network Slice Customer}
\newacronym{NSP}{NSP}{Network Slice Provider}
\newacronym{CSC}{CSC}{Communication Service Customer}
\newacronym{CSP}{CSP}{Communication Service Provider}
\newacronym{SST}{SST}{Slice/Service Type}
\newacronym{SD}{SD}{Slice Differentiator}
\newacronym{USRP}{USRP}{UE Router Selection Policy}
\newacronym{S-NSSAI}{S-NSSAI}{Single Network Slice Selection Assistance Information}
\newacronym{ONISTT}{ONISTT}{Open Net-centric Interoperability Standards for Training and Testing}
\newacronym{KB}{KB}{Knowledge Base}
\newacronym{NSI}{NSI}{Network Slice Instance}

\newacronym{VF}{VF}{Virtual Function}
\newacronym{VFC}{VFC}{Virtual Function Component}
\newacronym{CR}{CR}{Complex Resource}
\newacronym{PNF}{PNF}{Physical Network Function}
\newacronym{CP}{CP}{Connection Point}
\newacronym{VL}{VL}{Virtual Link}
\newacronym{SDC}{SDC}{Service Design and Creation}
\newacronym{ONAP}{ONAP}{Open Network Automation Platform}
\newacronym{VID}{VID}{Virtual Infrastructure Deployment}
\newacronym{VSP}{VSP}{Vendor Software Product}

\newacronym{WEF}{WEF}{Wireless Edge Factory}

\newacronym{DP}{DP}{Data Plane}
\newacronym{ECOMP}{ECOMP}{Enhanced Control Orchestration Management and Policy}

\newacronym{AAI}{AAI}{Active and Available Inventory}
\newacronym{SDNC}{SDNC}{Software Defined Network Controller}
\newacronym{SO}{SO}{Service Orchestrator}
\newacronym{APPC}{APPC}{Application Controller}
\newacronym{DCAE}{DCAE}{Data Collection Analytics and Events}
\newacronym{OOF}{OOF}{ONAP Optimization Framework}
\newacronym{OSS}{OSS}{Operation Support System}
\newacronym{BSS}{BSS}{Business Support System}
\newacronym{SOCKS}{SOCKS}{Secured Over Credential-based Keberos}
\newacronym{VVP}{VVP}{VNF Validation Program}
\newacronym{PDP}{PDP}{Policy Decision Point}
\newacronym{PEP}{PEP}{Policy Enforcement Point}
\newacronym{PCC}{PCC}{Policy Creation Component}
\newacronym{RU}{RU}{Remote Unit}

\newacronym{VLM}{VLM}{Vendor License Model}

\newacronym{I/Q}{I/Q}{in-phase / in-quadrature}
\newacronym{CUPS}{CUPS}{Control User Plane Separation}
\newacronym{OTT}{OTT}{Over The Top}
\newacronym{CAGR}{CAGR}{Compound Annual Growth Rate}
\newacronym{TowerCos}{TowerCos}{Tower Companies}